\documentclass[fleqn,10pt]{wlscirep}
\usepackage[utf8]{inputenc}
\usepackage[T1]{fontenc}\usepackage{physics}
\usepackage{graphicx}
\usepackage{float} 
\title{Non-Hermitian impurity problem}

\author[1,2]{E. T. Kokkinakis}
\author[1,2]{I. Komis}
\author[1,2, $\dagger$]{K. G. Makris}
\author[1,2, *]{E. N. Economou}

\affil[1]{Department of Physics, University of Crete, 70013 Heraklion, Greece}
\affil[2]{Institute of Electronic Structure and Laser (IESL), FORTH, 71110 Heraklion, Greece}
\affil[$\dagger$]{makris@physics.uoc.gr}
\affil[*]{economou@admin.forth.gr}

\begin{abstract}

{The problem of a single Hermitian impurity has long served as a cornerstone in condensed matter physics, offering fundamental insights into the mechanisms of Anderson localization. Yet, despite the increased interest in the spectral and localization properties of non-Hermitian lattices with defects, the non-Hermitian extension of the single impurity problem remains largely unexplored. In this work, we investigate the role of a single complex impurity in one-, two-, and three-dimensional infinite tight-binding lattices. Our study reveals a series of counterintuitive phenomena, including regions where localization vanishes and re-emerges as the impurity strength varies. Next, we study the corresponding finite-sized lattices, which are highly relevant to experimental realizations in readily accessible photonic platforms, revealing a variety of exotic features, such as scale-free localized states, exceptional points, and peculiar {cross-shaped} localized eigenstates, whose profiles deviate from the conventional exponential localization. This work paves the way for future studies on transport phenomena in non-Hermitian disordered lattices.}

\end{abstract}
\begin{document}

\flushbottom
\maketitle
%
%

\section*{Introduction}

Over the past two decades, non-Hermitian physics \cite{ElGanainy2018} has garnered significant attention as a robust framework for describing open systems. In particular, photonics has become the premier experimental platform \cite{Ruter2010} for physically realizing non-Hermitian Hamiltonians\cite{Moiseyev2011}, offering precise control over engineered distributions of gain and loss. Within this framework, extensive research efforts have explored, both theoretically and experimentally, concepts such as parity-time ($\mathcal{PT}$) symmetry \cite{Makris2008, ElGanainy2007, Musslimani2008, Guo2009, Regensburger2012, Hodaei2014, Konotop2016, Feng2017, Ozdemir2019}, exceptional points \cite{EP2, EP3}, non-Hermitian skin effect \cite{Hatano1996, lee_2016, yao_2018, harari_2018, bandres_2018, liu_2022, gao_2023} and non-Hermitian Anderson localization \cite{Longhi2019a, Chen2019, Longhi2019b, Chen2020, Tzortzakakis2020a, Tzortzakakis2020b, Tzortzakakis2021, Cao2021, Leventis2022, Bandyopadhyay2023, Kokkinakis2024, Datta2024, Zhai2024, He2024, Li2024, Samanta2025} - among  others - firmly establishing non-Hermitian photonics as a forefront of modern physics. More recently, attention has been devoted toward investigating non-Hermiticity in quantum many-body \cite{Fukui1998, Hamazaki2019, Liu2020, Zhai2020, Zhang2022_HN, Kawabata2022, Zhang2020_AAH, Ghosh2022, Gupta2024, Ganguly2024, Hu2025, DelgadoQuesada2025}, incoherent \cite{Longhi2024a, Longhi2024b, Longhi2024c, Kokkinakis2024b, Kokkinakis2025b}, and nonlinear \cite{Dobrykh2018, Hang2021, Ezawa2022, Zhu2022, longhi_2025, yuce_2025, wwang2025, Komis2023, Kokkinakis2025a} systems, underscoring the rich potential and future directions of this rapidly evolving field. 

In a seemingly unrelated area, defects have historically played an important role in solid-state physics by shaping the electronic properties of crystalline materials. Early seminal studies \cite{Koster1954, Koster1954_SIC} demonstrated that impurities and lattice vacancies can alter electron transport, modify scattering processes, and induce localization phenomena. By introducing localized potentials that trap electrons and modify band structures, defects give rise to phenomena such as defect-bound states and impurity bands—effects that have had a profound impact on semiconductor device physics \cite{Callaway1967, Callaway1967_LD, Bernholc1978}. Until lately, however, investigations of defect-related phenomena were limited to systems described by Hermitian Hamiltonians \cite{Economou2006}.

{Recently, defects have attracted renewed interest in non-Hermitian physics due to their profound influence on the properties of open systems. Research in this direction has primarily, though not exclusively \cite{Li2025, Yamamoto2025_AIM, Tombuloglu2025_Extreme}, focused on the role of impurities in lattices that are non-Hermitian in the absence of defects—such as Hatano–Nelson type \cite{Roccati2021, Zhao2025_Tentacle, Yang2025_Geometry} or $\mathcal{PT}$-symmetric \cite{Zhang2024, Liu2025, Stegmaier2021} models—with particular, emphasis on their topological characteristics \cite{Liang2025_Defect, LiuChen2019, Spring2024}. In parallel, scattering phenomena involving non-Hermitian defects have also drawn attention, including invisibility effects of driven impurities \cite{Longhi2017}, asymmetric transmission \cite{Burke2020}, and anomalous scattering \cite{Zhang2023}.}

{Regarding the localization properties of non-Hermitian systems with impurities, a particularly intriguing recent result relevant to such lattices, among other systems \cite{Yokomizo2021, Wang2022}, is the emergence of the so-called {scale-free localized} (SFL) states, which, unlike conventional exponentially localized states, exhibit localization characteristics that explicitly depend on the system size. Studies on these states have primarily focused on Hatano–Nelson-type lattices, either clean \cite{LiLeeGong2021, Liu2021, FuZhang2023, LiSunYangLi2024, Yuce2025} or in interplay with disordered and quasiperiodic potentials \cite{Molignini2023, Liu2024, Yilmaz2024, Y_Zhang2025}, while several works have explored their appearance in $\mathcal{PT}$-symmetric lattices containing imaginary defects \cite{Li2023, Guo2023}. Importantly, SFL states have been experimentally observed in electrical circuits \cite{LiLeeGong2021, Guo2024, Xie2024, Wang2025} and superconducting platforms \cite{Zhang2025}.}

{However, despite the existence of numerous studies related to non-Hermitian systems with defects, the fundamental problem of localization for a single {non-Hermitian} impurity in a {Hermitian} tight-binding lattices remains unresolved—specifically, whether and under what conditions such a defect can induce a localized state, both in the thermodynamic limit and in finite-size lattices. This question lies at the core of understanding localization phenomena in non-Hermitian disordered systems and forms the focus of the present work.}

{In this work, we investigate the impact of a single complex impurity on the spectral and localization properties of one-dimensional (1D), two-dimensional (2D) square, and three-dimensional (3D) cubic tight-binding lattices. To elucidate the fundamental mechanisms underlying impurity-induced localization, we first address the single complex impurity problem in the thermodynamic limit, revealing a wealth of phenomena far more intricate than those found in the well-studied Hermitian case. We then examine the spectral and localization characteristics of the corresponding finite lattices, where the fingerprints of the infinite system persist alongside additional effects unique to finite sizes. The results for infinite lattices are of direct relevance to condensed-matter physics, as they establish the non-Hermitian counterpart of the single-impurity problem, while the findings for finite lattices are particularly pertinent to non-Hermitian photonics, where such systems can be readily realized experimentally, e.g. in waveguide lattices, cavity arrays or photonic synthetic-dimension mesh-lattices \cite{Christodoulides2003, liu_2022, wwang2025}.}
\section*{Results}
Let us begin by considering the 1D, 2D square, and 3D cubic  Hermitian tight-binding lattices with uniform nearest-neighbor hopping terms (set to \(c \equiv 1\)), under the presence of a single non-Hermitian defect. In all cases, the defect is introduced via an on-site impurity potential $\gamma$ at lattice site $\ket{{\mathbf{m}}}$, where \(\gamma^* \neq \gamma\). Therefore, the non-Hermitian Hamiltonian describing the 1D lattice with \(N\) sites is given by
\begin{equation}
    \label{hamiltonian_1d}
    H_{1D} = \gamma\,\ket{m}\bra{m} +\sum_{n=1}^{N-1} \left( \ket{n}\bra{n+1} + \ket{n+1}\bra{n} \right) 
\end{equation}
Throughout this work, we denote the unperturbed 1D Hamiltonian $H_{1D}(\gamma=0)$ as \(H_0\). To extend the model to higher dimensions, we express the Hamiltonians for the corresponding 2D (\(N \times N\) sites) and 3D (\(N \times N \times N\) sites) lattices as
\begin{subequations}
\label{hamiltonians}
\begin{align}
    H_{2D} &= H_0 \otimes \mathbb{I}_{N} + \mathbb{I}_{N} \otimes H_0 + \gamma\,\ket{\mathbf{m}}\bra{\mathbf{m}}, \label{hamiltonian_2d} \\
    H_{3D} &= H_0 \otimes \mathbb{I}_{N} \otimes \mathbb{I}_{N}
            + \mathbb{I}_{N} \otimes H_0 \otimes \mathbb{I}_{N}
            + \mathbb{I}_{N} \otimes \mathbb{I}_{N} \otimes H_0
            + \gamma\,\ket{\mathbf{m}}\bra{\mathbf{m}}. \label{hamiltonian_3d}
\end{align}
\end{subequations}
Here, \(\mathbb{I}_N\) denotes the \(N\)-dimensional identity matrix, and lattice site indices are represented as \(\ket{\mathbf{n}} \equiv (n_x,n_y)\) in 2D and \(\ket{\mathbf{n}} \equiv (n_x,n_y,n_z)\) in 3D.

We organize our results as follows: In the first part of this work, we focus on the case of infinite-sized lattices (\(N \to \infty\)), which is related to condensed matter physics, where system sizes typically are \(N \sim 10^8\) or larger. We denote this thermodynamic limit using a tilde notation: \(\tilde{H}_d \equiv H_d(N \to \infty)\), with \(d \in \{\text{1D, 2D, 3D}\}\) denoting the dimensionality. In the second part, we shift our attention to finite-sized lattices, which are more relevant to experimental realizations, particularly within the field of non-Hermitian photonics. For reference and comparison, we begin each subsection by briefly revisiting the well-established case of a real defect (\(\gamma = V >0\)), before proceeding to the non-Hermitian scenarios involving a purely imaginary (\(\gamma = iV \) with $V>0$) or, in general, a complex (\(\gamma = V_1+iV_2 \) with $V_1,V_2>0$) impurity.

\subsection*{Infinite-sized lattice}
The problem of a single Hermitian impurity has been extensively studied in solid-state physics \cite{Economou2006}. In particular, the Green's function formalism provides a direct method for determining whether an on-site perturbation in an infinite lattice can induce a bound (localized) state, for both the Hermitian (\(\gamma^* = \gamma\)) and the non-Hermitian (\(\gamma^* \neq \gamma\)) impurity cases. The discrete eigenvalue \(\epsilon_b = a + ib \) (with $b=0$ for the Hermitian and $b\neq 0$ for the non-Hermitian case, respectively) that corresponds to a bound state of the perturbed infinite-lattice Hamiltonian \(\tilde{H}_d(\gamma \neq 0)\) is determined by the pole of its Green's function, \(G(\epsilon) \equiv (\epsilon - \tilde{H}_d)^{-1}\), where \(\epsilon = x + iy \in \mathbb{C}\), generally. This pole satisfies the condition
\begin{equation}
\label{pole}
1 - \gamma G_0(\epsilon_b) = 0,
\end{equation}
for \(\epsilon = \epsilon_b\) lying {outside} the eigenvalue band of the unperturbed Hamiltonian \(\tilde{H}_d(\gamma = 0)\). Here, \(G_0(\epsilon)\) denotes the diagonal element of the unperturbed Green's function, defined as \(G_0(\epsilon) \equiv G_0(\mathbf{n}, \mathbf{n}; \epsilon) = \bra{\mathbf{n}}(\epsilon- \tilde{H}_d(\gamma = 0))^{-1} \ket{\mathbf{n}}\), which depends on the dimensionality of the system and is given, for $\text{Re}(\epsilon)>0$ and $\text{Im}(\epsilon)>0$ outside the unperturbed bands, explicitly by
\begin{equation}
G_0(\epsilon) =
\begin{cases}
\displaystyle \frac{1}{\sqrt{\epsilon^2 - 4}}, & \text{for 1D,} \quad \epsilon \notin [0, 2], \\[1em]
\displaystyle \frac{2}{\pi \epsilon}\,\mathbb{K}\left(\frac{4}{\epsilon}\right), & \text{for 2D,} \quad \epsilon \notin [0, 4], \\[1em]
\displaystyle \frac{1}{2\pi^2} \int_0^{\pi} t\, \mathbb{K}(t)\, d\phi, & \text{for 3D,} \quad \epsilon \notin [0, 6],
\end{cases}
\end{equation}
where \(\mathbb{K}(u) \equiv \frac{1}{2} \int_0^{\pi} \frac{d\theta}{\sqrt{1 - u^2 \cos^2 \theta}}\) is the complete elliptic integral of the first kind, and \(t \equiv \frac{4}{\epsilon - 2\cos{\phi}}\).
The existence of bound states for a given defect strength \(\gamma\) depends on whether Eq.~\eqref{pole} admits a solution. 

\subsubsection*{Real impurity}
For a Hermitian defect (\(\gamma = V >0\)), the condition \(\Im(G_0(\epsilon)) = 0\) must be satisfied, which holds for values \(\epsilon = x \in \mathbb{R}\) lying outside the spectrum of \(\tilde{H}_d(\gamma = 0)\). Thus, Eq. \eqref{pole} simplifies to \(\Re(G_0(x)) = 1/V\). Of course, if \(\Re(G_0(x))\) has a finite maximum, a threshold for $V$ is required for this equation to be satisfied.  Therefore, the absence or presence of singularities in \(\Re(G_0(x))\) determines whether a threshold defect strength is required or not, respectively, for bound-state formation. 

As it is well established, in 1D, $\Re(G_0(x))$ shows a square-root singularity as \(x \to 2^+\), while in 2D, a logarithmic singularity occurs, respectively, as \(x \to 4^+\); in both cases  bound states are allowed to form for arbitrarily small \(V\), noting that $\Re(G_0(x))$ is monotonically decreasing for $x$ outside the unperturbed band. In contrast, in 3D, as \(x \to 6^+\), \(\Re (G_0(x))\) exhibits only a derivative discontinuity and remains finite, approaching \(\Re (G_0(6^+)) \to 0.252\), which is its maximum for $|x|>6$. As a result, a finite threshold \(V_{\text{th}} \approx 3.95\) is required for bound-state formation for the 3D lattice.
As we will see, these properties change significantly when the defect is non-Hermitian. 

\subsubsection*{Imaginary impurity}
For a purely imaginary defect $\gamma = iV$ with $V > 0$, Eq.~\eqref{pole} requires $\Re\left(G_0(\epsilon\right) = 0$. This condition is satisfied along the imaginary axis ($\epsilon = iy$) in all dimensions, simplifying Eq.~\eqref{pole} to $\Im\left(G_0(iy)\right) = -1/V$. Analogous to the real-impurity case, the existence of a bound state depends on the behavior of $\Im(G_0(iy))$; if it has a finite minimum, a threshold value of $V$ must be exceeded, while if it diverges, the condition is always satisfied for some $y > 0$.

Particularly, in 1D, the function \(\Im\left(G_0(iy)\right)\) approaches \(-1/2\) as \(y \to 0^{+}\) and does not exhibit any singularities, implying the existence of a finite threshold for bound-state formation. Solving Eq. \eqref{pole} yields a bound-state eigenvalue \(\epsilon_b = i\sqrt{V^2 - 4}\) , which lies outside the interval \([0, 2]\), provided that \(V \geq 2 \equiv V_{\text{th}}\). This result is in sharp contrast to the Hermitian case, where bound states form even for arbitrarily small defect strengths in 1D.
In 2D, the imaginary part of the Green's function along the imaginary axis, \(\Im\left(G_0(iy)\right)\), exhibits a logarithmic singularity as \(y \to 0^{+}\), given by $\Im\left(G_0(iy)\right) \to -\frac{1}{2\pi} \ln\left(\frac{16}{y}\right)$.
Thus, it does not possess a finite minimum, and the condition \(\Im\left(G_0(iy)\right) = -1/V\) always admits a solution \(\epsilon_b\), even for infinitesimally small \(V\), similarly to the Hermitian impurity case.
Conversely, for the 3D lattice, the function \(\Im\left(G_0(iy)\right)\) attains a finite minimum as \(y \to 0^{+}\), approximately equal to \(-0.448\). As a result, a finite threshold \(V_{\text{th}} \approx 2.23\) is required for the formation of a bound state—substantially lower than the corresponding threshold for the Hermitian defect. {Let us here note that the quantity $\Im\left(G_0(iy)\right)$ for $y\to 0^{+}$ is proportional to the unperturbed lattice's density of states $\rho(\epsilon)$ at the band center $\epsilon=0$, i.e., $\lim_{y\to0^{+}} \Im\left(G_0(0+iy)\right)=-\pi \rho(0)$ \cite{Economou2006}. }

\subsubsection*{Complex impurity}
As evident from the previous paragraph, a purely imaginary impurity exhibits markedly different behavior regarding bound-state formation compared to its Hermitian counterpart. However, the problem becomes significantly more intricate in the case of a complex impurity, i.e., \(\gamma = V_1 + iV_2\) with \(V_1, V_2 >0\). In this general scenario, Eq.~\eqref{pole} can be equivalently expressed through the following  conditions:
\begin{subequations}
\begin{align}
\Re\bigl(G_0(\epsilon)\bigr) 
  &= \frac{V_1}{V_1^2 + V_2^2} 
  \label{cond_real} \\ 
\Im\bigl(G_0(\epsilon)\bigr) 
  &= -\frac{V_2}{V_1^2 + V_2^2} 
  \label{cond_imag}
\end{align}
\end{subequations}

For a bound state to exist, these two equations must admit a {common} solution. Let \(L_1\) and \(L_2\) denote the curves satisfying these two conditions, respectively. Then, bound-state formation becomes impossible in two cases: (a) if either equation has no solution—i.e., if \(L_1\) or \(L_2\) does not exist—and (b) if the two curves \(L_1\) and \(L_2\) do not intersect. 
\begin{figure}{H}
    \centering
    \includegraphics[width=0.8\textwidth]{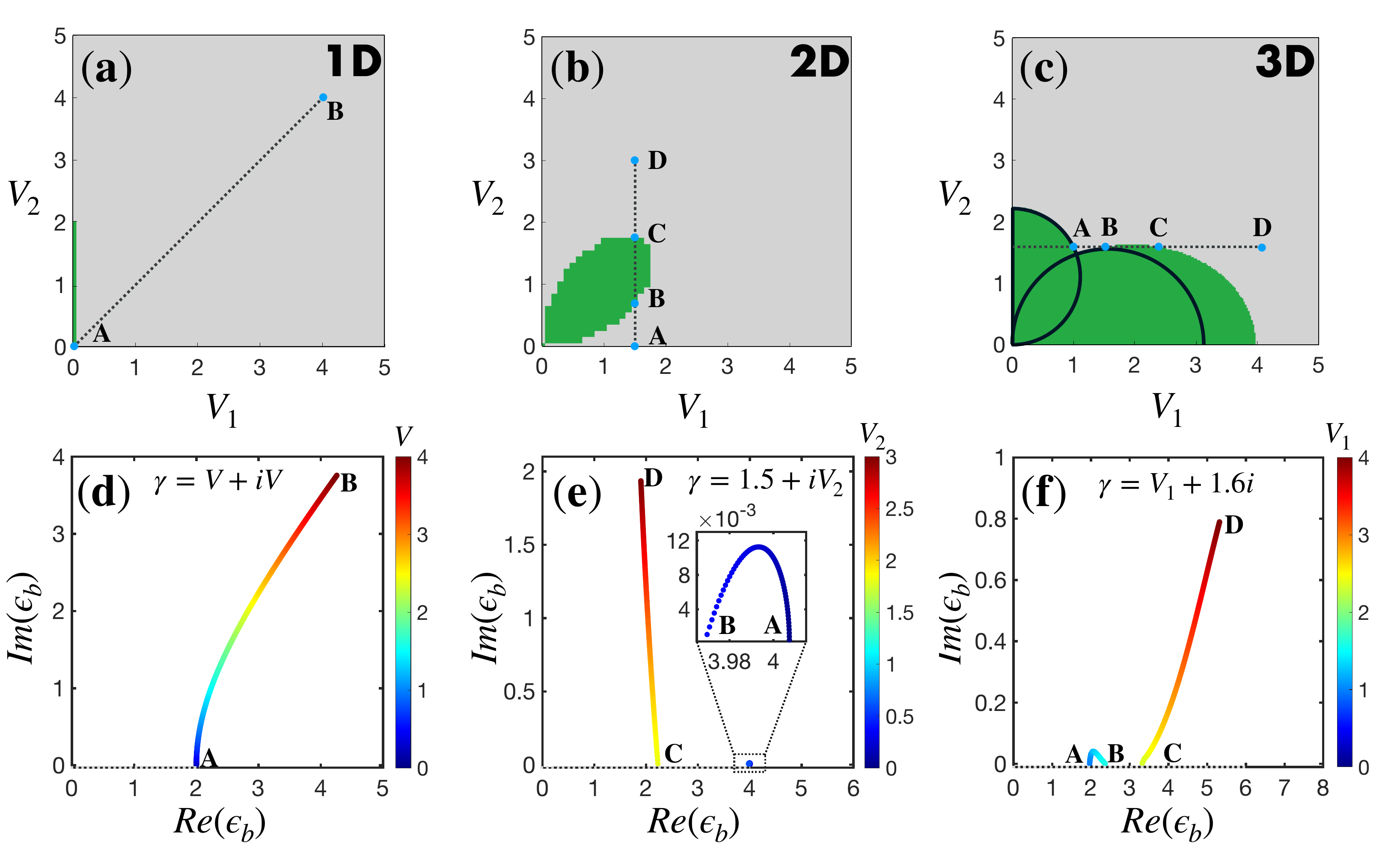}
    \caption{\textbf{Bound-state existence maps and trajectories for a complex impurity \(\gamma = V_1 + iV_2\) as a function of \(V_1\) and \(V_2\).}
(a) Map of bound-state existence corresponding to 1D infinite-sized lattice, where regions colored gray indicate bound-state existence, while the respective green-colored regions indicate absence. (b) Similarly with (a), for 2D infinite-sized square lattice. (c) Similarly with (a), for 3D infinite-sized cubic lattice. (d) Trajectory of the bound-state eigenvalue \(\epsilon_b\) in the complex plane, for the 1D lattice with impurity \(\gamma = V + iV\) as a function of \(V \in [0,4]\) (shown in the color bar). 
(e) Similarly with (d), with impurity \(\gamma = 1.5 + iV_2\) as a function of \(V_2 \in [0,3]\). (f) Similarly with (d), with impurity \(\gamma = V_1 + 1.6i\) as a function of \(V_1 \in [0,4]\). In each of (d)–(f), the positive part of the unperturbed eigenvalue band is shown dotted. Corresponding points in parameter space of (a)–(c) and eigenvalue trajectories of (d)–(f) are marked by bold capital letters.
} 

    \label{map}
\end{figure}
We begin our analysis with the 1D case. As established earlier, a purely imaginary defect (\(V_1 = 0\)) requires a finite value of defect strength, \(V_2 \geq 2\), for bound-state formation. However, the situation changes dramatically with the introduction of an infinitesimal real component (\(V_1 > 0\)). In this case, both conditions—Eqs.~\eqref{cond_real} and~\eqref{cond_imag}—are satisfied, as the real and imaginary parts of \(G_0(\epsilon)\) both exhibit square-root singularities near the band edge \(\epsilon = 2\). The proximity of these singularities ensures that the corresponding curves \(L_1\) and \(L_2\), which always exist, intersect for any \(V_1 > 0\), regardless of how small \(V_2\) is. Therefore, for the 1D lattice, bound states fail to form by a complex defect, {only} in the case \(V_1 = 0\) and \(V_2 < 2\), as discussed in the previous paragraph. This behavior is illustrated schematically in Fig.~1(a), which shows the region in the \(V_1\text{/}V_2\) plane (gray area) where bound states are supported. Additionally, Fig.~1(d) demonstrates the trajectory of the bound-state eigenvalue \(\epsilon_b\) in the complex plane for a specific representative case, \(\gamma = V + iV\). For \(V \to 0\), the eigenvalue approaches the band edge \(\epsilon = 2\), where both \(\Re(G_0(\epsilon))\) and \(\Im(G_0(\epsilon))\) diverge, enabling the pole condition to be satisfied even for infinitesimally small \(V\).  In the opposite limit \(V \gg 1\), the eigenvalue $\epsilon_b$ asymptotically approaches \(\gamma = V+iV\).

The situation becomes significantly more intricate in the 2D case. Although both \(\Re(G_0(\epsilon))\) and \(\Im(G_0(\epsilon))\) exhibit singularities—ensuring the existence of the curves \(L_1\) and \(L_2\) for all \(\epsilon \in \mathbb{C}\)—these singularities occur in different regions of the complex plane: the singularity in \(\Re(G_0(\epsilon))\) arises near \(\epsilon = 4\), while that in \(\Im(G_0(\epsilon))\) appears near \(\epsilon = 0\). As a result, for small nonzero values of both \(V_1\) and \(V_2\), the curves \(L_1\) and \(L_2\) are spatially separated and do not intersect, making bound-state formation impossible. This leads to the emergence of a {hole} in the \(V_1\text{/}V_2\) parameter space where no bound state is supported, as shown in the green area of Fig.~1(b).
This outcome {}{in the 2D case} is counterintuitive; although an infinitesimal defect strength—either purely real or purely imaginary—is sufficient to induce a bound state, their simultaneous presence can fail to do so within a finite region of the \(V_1\text{/}V_2\) space. As a representative example, Fig.~1(e) tracks the trajectory of the eigenvalue \(\epsilon_b\) when the real part of the defect's strength is fixed at \(V_1 = 1.5\), while \(V_2\) increases from 0 to 3. The eigenvalue begins at \(\epsilon_b \approx 4.007\) and follows an arc-like trajectory near \(\epsilon = 4\), developing a non-zero imaginary part. It then collapses onto the real axis for \(V_2 \approx 0.7\), indicating inability to form a bound state. The eigenvalue $\epsilon_b$ re-emerges from the interior of the unperturbed band at \(x \approx 2.22\) for \(V_2 \gtrsim 1.8\), and subsequently asymptotically approaches \(\gamma = 1.5 + iV_2\) for \(V_2 \gg 1\).

In the 3D case, bound-state formation is hindered by the fact that neither \(\Re(G_0(\epsilon))\) nor \(\Im(G_0(\epsilon))\) exhibit singularities. Instead, they attain finite extrema: \(\Re(G_0(\epsilon))\) reaches a maximum of approximately 0.45, while \(\Im(G_0(\epsilon))\) attains a minimum of about \(-0.32\). Consequently, it follows directly from Eqs.~\eqref{cond_real} and~\eqref{cond_imag} that the curves \(L_1\) and \(L_2\) do not exist within certain regions of the parameter space. Specifically, \(L_1\) is absent inside the semicircular region defined by \((V_1 - 1.562)^2 + V_2^2 \leq 1.562^2\), and \(L_2\) does not exist within \((V_2 - 1.15)^2 + V_1^2 \leq 1.15^2\). Thus, in both regions, whose boundaries are shown in Fig. 1(c), the conditions for bound-state formation are not met. Beyond these exclusion zones, there is also a range of values in the \(V_1\text{/}V_2\) plane for which the curves \(L_1\) and \(L_2\) exist yet do not intersect, once again preventing the emergence of a bound state. This complex behavior is illustrated in Fig.~1(c), where the nontrivial structure of the region where bound-state formation is forbidden (green area) is shown.  This leads to counterintuitive phenomena, including the formation, disappearance, and reappearance of bound states. A representative example is shown in Fig.~1(f), which tracks the eigenvalue trajectory for a defect strength with fixed imaginary part \(V_2 = 1.6\), and rising real part \(V_1\) from 0 to 4. Initially, a threshold at \(V_1 \approx 1\) exists, beyond which a bound state emerges near \(\epsilon \approx 2\), close to the maximum of \(\Re(G_0(\epsilon))\). As \(V_1\) increases further, the eigenvalue follows an arc-like path before collapsing into the unperturbed band at \(\epsilon_b \approx 2.33\) for \(V_1 \approx 1.6\). The bound state then re-emerges with \(\epsilon_b \approx 3.32\) when \(V_1 > 2.3\). For much larger \(V_1\) values, the eigenvalue asymptotically approaches \(\gamma = V_1 + 1.6i\), as expected.

In summary, the existence of bound states in infinite tight-binding lattices depends strongly on the defect’s complex strength \(\gamma = V_1 + iV_2\) and the lattice's dimensionality. As it is well known, a purely real impurity (\(V_2 = 0\)) yields bound states for arbitrarily small \(V_1\) in 1D and 2D, but requires \(V_1 \approx 3.95\) in 3D. On the other hand, we showed that a purely imaginary impurity (\(V_1 = 0\)) allows the existence of bound states at any strength in 2D, but demands \(V_2 \geq 2\) in 1D and \(V_2 \gtrsim 2.23\) in 3D. With both real and imaginary parts nonzero, in 1D always a bound state is formed as long as \(V_1 > 0\), whereas in 2D a finite {hole} in the \(V_1\text{/}V_2\) plane exists, where no bound state is allowed. In 3D, there are semicircular exclusion zones and additional parameter regions wherein the pole condition Eq. \eqref{pole} is not satisfied.
The aforementioned results for infinite-sized lattices of all dimensions (1D, 2D or 3D), regarding the regions \(V_1\text{/}V_2\) where bound state formation is prohibited or allowed, are shown in green and gray color, respectively, in the top row of Fig. 1.

\subsection*{Finite-sized lattices}
Having established a comprehensive presentation of the infinite-lattice problem, we now turn to the case of finite-sized lattices. This regime is particularly relevant to experimental realizations in readily accessible photonic systems, such as waveguide arrays, synthetic mesh lattices, and optical cavity arrays, where system sizes are inherently finite.

As will become evident in what follows, the properties of the infinite lattice leave clear fingerprints on the corresponding finite-sized systems with a single local impurity. However, in addition to capturing signatures of the thermodynamic limit, here we aim to study the spectral and localization features of {all} eigenvalues and eigenmodes of the finite-sized system, extending beyond the eigenstate corresponding to the bound state. In particular, we focus on the behavior of eigenvalues in the complex plane for various impurity strengths and explore their relation to the spatial profiles of the associated eigenmodes. To this end, we first analyze in detail the specific scenario of a purely imaginary defect. This representative case encapsulates all qualitative phenomena emerging in lattices with a single complex impurity, while being also convenient to study due to the symmetry of its eigenspectrum with respect to the imaginary axis, as we will discuss later.

\subsubsection*{Real impurity}
For reference, we briefly discuss the spectral characteristics of finite-sized tight-binding lattices with a single real impurity \(\gamma = V>0\). The eigenvalue equation for the Hamiltonian \(H_{d}\) reads
$H_{d}\ket{u_j} = \epsilon_j\ket{u_j}$,
where \(\ket{u_j}\) is the eigenstate and \(\epsilon_j\) is the corresponding eigenvalue, with \(d \in \{\text{1D, 2D, 3D}\}\). The amplitude of \(\ket{u_j}\) at lattice site \(\mathbf{n}\) is denoted as \(u_{j,\mathbf{n}} \equiv \bra{\mathbf{n}}\ket{u_j}\).

In the 1D case under open boundary conditions (OBC), defined by \(\ket{0} = \ket{N+1} = 0\), the unperturbed Hamiltonian \(H_0\) has eigenvalues \(\epsilon_j^{(0)} = 2\cos\left( \frac{j\pi}{N+1} \right)\), with corresponding eigenstates \(u_{j,n}^{(0)} \propto \sin\left( \frac{n\pi j}{N+1} \right)\). Any unperturbed eigenstate that vanishes at the defect site \(m\), i.e., \(u_{r,m}^{(0)} = 0\), remains unaffected for any value of the defect strength \(V\), i.e., \(\epsilon_r = \epsilon_r^{(0)}\) and \(u_{r,m} = u_{r,m}^{(0)}\). The number of such unaffected eigenstates depends on both the total system size \(N\) and the defect position \(m\).
For periodic boundary conditions (PBC), defined by \(\ket{0} = \ket{N}\) and \(\ket{1} = \ket{N+1}\), the unperturbed spectrum becomes \(\epsilon_j^{(0)} = 2\cos\left( \frac{2j\pi}{N} \right)\), and eigenvalue degeneracies appear. One can show that the number of eigenvalues unaffected by the defect is \(\frac{N+1}{2}\) for odd \(N\), and \(\frac{N}{2} + 1\) for even \(N\).
In higher dimensions (2D and 3D), the unperturbed spectra include degeneracies, and the number of defect-unaffected eigenstates—under either OBC or PBC—depends strongly on the system size.

Importantly, the choice of boundary conditions significantly influences the spectral response to the defect, making PBC a choice that captures the characteristics of the infinite lattice for smaller lattices sizes than OBC do. Under PBC, the largest unperturbed eigenvalue exactly matches the band edge of the infinite lattice, even for small \(N\), across all dimensions. Moreover, an infinitesimal defect strength \(V\) is sufficient to induce a localized state in 1D and 2D. In 3D, however, inducing a localized state requires a non-zero value of \(V\), which, as \(N\) increases, converges to the infite-sized lattice's threshold \(V_{\text{th}} \approx 3.95\). In contrast, under OBC, the largest unperturbed eigenvalue \(\epsilon_j^{(0)}\) is always smaller than the infinite-lattice band edge, though it gradually approaches this limit as \(N\gg 1\). Also, OBC introduce substantial finite-size effects, particularly in small lattices and thus, in order to accurately capture the signatures of the infinite system in this case, one requires typically a sufficiently large lattice and a defect positioned far from the boundaries.
Nevertheless, given their direct relevance to experimental implementations—such as photonic waveguide arrays, where open boundaries inherently exist—we will focus on OBC in the following subsections.

\subsubsection*{Imaginary impurity}

We now consider the case of a purely imaginary impurity, \(\gamma = iV\), with \(V >0\). The resulting Hamiltonian $H_d$ is non-Hermitian, $H_d\neq H_d^{\dagger}$. The respective {right} eigenvalue equation \cite{Moiseyev2011} reads $H_{d}\ket{u_j} = \epsilon_j\ket{u_j},$
where \(\ket{u_j}\) is the {right} eigenstate and \(\epsilon_j = a_j + i b_j\) is {}{the corresponding eigenvalue (which in general is expected to be complex)}. 
It can be shown [see Supplementary Note 1] that the spectrum \(\{\epsilon_j\}\) of \(H_{d}\) is symmetric with respect to the imaginary axis, i.e., if \(E\) is an eigenvalue of \(H_{d}\) (\(H_{d}\ket{\psi}=E\ket{\psi}\)), then \(-E^{*}\) is also an eigenvalue (\(H_{d}\ket{\phi}=-E^{*}\ket{\phi}\)). Furthermore, the corresponding right eigenstates \(\ket{\psi}\) and \(\ket{\phi}\) have equal amplitudes, namely, $\abs{\psi_{\mathbf{n}}} \equiv \abs{\bra{{\mathbf{n}}}\ket{\psi}} = \abs{\phi_{\mathbf{n}}} \equiv \abs{\bra{{\mathbf{n}}}\ket{\phi}}$,
for all lattice sites \(\ket{{\mathbf{n}}}\). 

\textbf{1D lattice:} We will begin our analysis with the 1D lattice case, under OBC. Similarly to the real-defect case discussed above, eigenvalues \(\epsilon_r\) associated with an unperturbed eigenstate that vanishes at the defect site, i.e., \(u_{r,m}^{(0)} = 0\), remain unaffected for any defect strength \(V\). In such cases, \(\epsilon_r = \epsilon_r^{(0)}\) and \(u_{r,m} = u_{r,m}^{(0)}\), while all other eigenvalues \(\epsilon_j\) are modified by the presence of the defect and acquire a nonzero imaginary part \(\Im(\epsilon_j) = b_j \neq 0\), even for infinitesimal \(V\). In what follows, we examine two representative lattice configurations. In the first configuration, the defect is placed at the center of an odd-sized lattice (\(N = 2M + 1\), with \(M \in \mathbb{N}\)), i.e., at site \(m= M + 1\). This setup preserves mirror symmetry with respect to the lattice's center. In this case, exactly \((N - 1)/2\) eigenvalues remain purely real, while the remaining \((N + 1)/2\) eigenvalues acquire nonzero imaginary parts. In the second configuration, the defect is located at site \(m = M\) of an even-sized lattice (\(N = 2M\)); here, in general, all eigenvalues become complex. Although this configuration lacks perfect mirror symmetry, the defect is positioned far enough from the boundaries to minimize edge effects. In what follows, for convenience, we index the eigenvalues \(\{\epsilon_j\}\) in descending order of their imaginary parts, such that \(\Im(\epsilon_1) \geq \Im(\epsilon_2) \geq \cdots \geq \Im(\epsilon_N)\), where $N$ is the total number of lattice sites.

To illustrate the aforementioned characteristics, Fig.~2 (top row) presents pertinent eigenvalue spectra for both configurations: a mirror symmetric lattice with \(N = 51\) and the defect placed at $m=26$ [Figs.~2(a) and 2(c)] and a non-mirror symmetric lattice with \(N = 50\) and the defect placed at $m=25$ [Figs.~2(b) and 2(d)]. {}{We observe that, in sharp contrast to the infinite lattice—where at most one eigenvalue is pulled out of the band—a single impurity in a finite lattice drastically alters the eigenvalue spectrum.} For a defect strength \(V = 1\) [Figs.~2(a)-(b)], no eigenvalue clearly separates from the rest of the spectrum, and the two largest imaginary parts remain equal: \(\Im(\epsilon_1) = \Im(\epsilon_2)\). However, for a stronger defect [\(V = 2.1\), Figs.~2(c), 2(d)], a single eigenvalue emerges with a distinctly larger imaginary part, becoming isolated from the rest of the spectrum. {}{This eigenvalue corresponds to the single bound state that occurs in the infinite‐lattice size limit, while all the others return to the real axis in that limit, as we will see in what follows}.

\begin{figure*}
    \centering
    \includegraphics[width=0.9\textwidth]{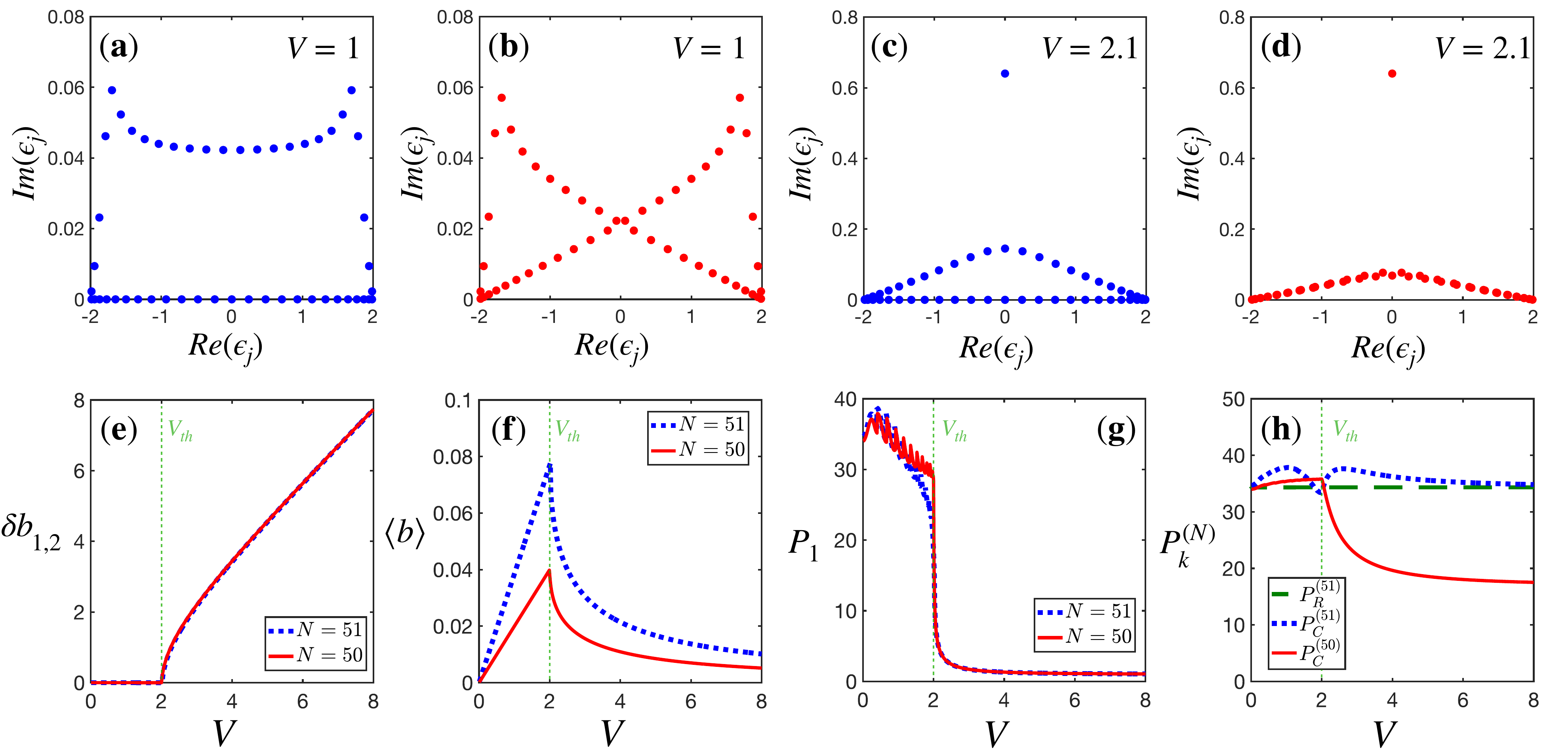}
    \caption{\textbf{Spectral properties of 1D finite-sized lattices with a single imaginary impurity.} (a)–(d) Eigenvalue spectra of 1D tight-binding {}{finite-sized} lattices ($N$ sites) with a single imaginary defect of strength $V$ located at site $m$. (a)/(c) Mirror-symmetric case ($N = 51$, $m = 26$) for $V = 1$ and $V = 2.1$, respectively; (b)/(d) non-mirror-symmetric case ($N = 50$, $m = 25$) for the same values of $V$. {}{Notice that all eigenvalues in (b) and (d), and $(N+1)/2$ of them in (a) and (c), were affected by the presence of the single defect by acquiring an imaginary part.} 
    (e) Difference \(\delta b_{1,2} = \Im(\epsilon_1) - \Im(\epsilon_2)\) as a function of $V$.
(f) Mean imaginary part \(\langle b \rangle\) of non-real eigenvalues, excluding \(b_1\) for \(V > V_{\text{th}}\), as a function of $V$. 
(g) Participation ratio \(P_1\) of the eigenstate with the largest imaginary part, as a function of $V$.  
(h) Average participation ratios of eigenstates with purely real (\(P_R\), green) and complex (\(P_C\)) eigenvalues (excluding \(\ket{u_1}\) for \(V > V_{\text{th}}\)), as a function of $V$.  
In (e)–(h), blue and red curves correspond to the mirror- and non-mirror-symmetric lattice cases, respectively., while the green vertical line marks the threshold value $V_{\text{th}}$ for formation of a localized state.
}

    \label{fig:Schematic}
\end{figure*}
\begin{figure}
    \centering
    \includegraphics[width=0.44\textwidth]{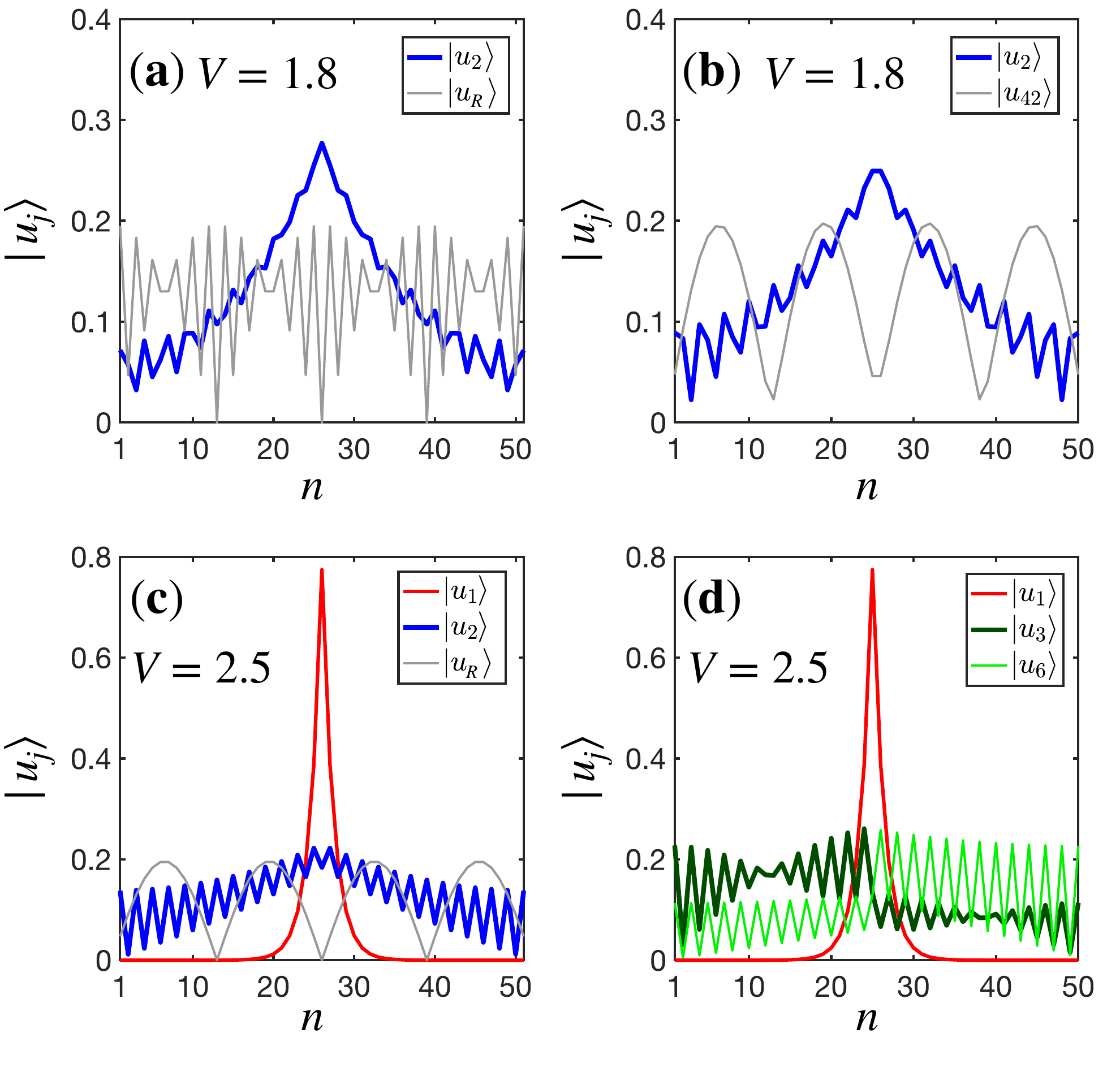}
    \caption{\textbf{Amplitudes \(|u_{j,n}|\) of representative right eigenmodes \(\ket{u_j}\) for 1D {finite-sized} lattices with a single imaginary impurity. }(a)/(c) correspond to the mirror-symmetric case (\(N=51\), \(m=26\)), while (b)/(d) pertain to the non-mirror symmetric configuration (\(N=50\), \(m=25\)), where $N$ is the number of lattice sites and $m$ the site of the impurity. For \(V=1.8<V_{\text{th}}\) (a)-(b), extended (gray) and scale-free localized (SFL) modes (blue) are observed; for \(V=2.5>V_{\text{th}}\) (c)-(d), an exponentially localized (red) state emerges. In (c) the SFL modes persist, whereas in (d) they become scattering states (green). In particular, the extended real-eigenvalue eigenmodes (gray) $\ket{u_R}$ of (a) and (c) correspond to ${u_{R,n}}\propto \sin(\frac{7\pi n}{13})$ and ${u_{R,n}}\propto \sin(\frac{\pi n}{13})$, respectively.}

    \label{fig:Schematic}
\end{figure}

As shown in Fig.~2(e), the dependence of \(\delta b_{1,2} \equiv \Im(\epsilon_1) - \Im(\epsilon_2)\) on the defect strength \(V\) reveals the emergence of a critical threshold at \(V_{\text{th}} = 2\), a clear fingerprint of the bound-state formation discussed previously for the infinite-sized lattice. For \(V < 2\), the two eigenvalues with the largest imaginary parts remain symmetric with respect to the imaginary axis. It must be noted that in the case of an even-sized lattice with the defect placed at \(m = M\), all eigenvalues form pairs that coalesce at \(V = V_{\text{th}} = 2\), corresponding to {pairs of exceptional points} \cite{Burke2020}. In contrast, for \(V > 2\), a single eigenvalue detaches from the rest of the spectrum, acquiring a larger imaginary part that increases monotonically with \(V\).

To further examine the spectral properties of the system, we define \(\langle b \rangle\) as the mean imaginary part of all complex eigenvalues, excluding the isolated eigenvalue that emerges for \(V > 2\). As shown in Fig.~2(f), the imaginary parts of the complex eigenvalues increase as \(V\) approaches the threshold \(V_{\text{th}} = 2\), but beyond this value, the average \(\langle b \rangle\) decreases monotonically. Moreover, as detailed in the Supplementary Note 2, \(\langle b \rangle \to 0\) with increasing system size \(N\) (as $\langle b \rangle\propto 1/N$) regardless of the value of \(V\). In contrast, the imaginary part \(b_1 = \Im(\epsilon_1)\) of the isolated eigenvalue for \(V \geq 2\) remains constant with increasing $N$, indicating that in the thermodynamic limit \(N \to \infty\), only a single eigenvalue remains separated from the rest of the spectrum, as we have already studied in the first part of this work.
 
This spectral behavior is closely linked to spatial localization. As previously mentioned, the threshold \(V_{\text{th}} = 2\) marks not only the spectral detachment of a single eigenvalue but also the emergence of a localized eigenstate. A convenient measure to quantify the degree of eigenmode localization is the so-called participation ratio. Let \(d\in\{1,2,3\}\) and consider a \(d\)-dimensional lattice with \(N\) sites along each direction (total \(M = N^d\) sites).  For a wavefunction \(\ket{\xi}\) with components
$ \xi_{n_1,\dots,n_d}
\;=\;
\bigl\langle n_1,\dots,n_d \,\big|\;\xi \bigr\rangle
\quad
(n_i=1,\dots,N),$
the {participation ratio} is defined by
\begin{equation}
   P \;\equiv\;
\frac{\displaystyle
  \Bigl(\sum_{n_1=1}^N\!\cdots\!\sum_{n_d=1}^N \bigl|\xi_{n_1,\dots,n_d}\bigr|^2\Bigr)^{2}}
{\displaystyle
  \sum_{n_1=1}^N\!\cdots\!\sum_{n_d=1}^N \bigl|\xi_{n_1,\dots,n_d}\bigr|^4}
\;. 
\end{equation}
For a fully extended state, \(\xi_{n_1,\dots,n_d} = 1/\sqrt{N^d}\), hence \(P = N^d\).
For a completely localized state, \(\xi_{n_1,\dots,n_d} = \delta_{n_1,m_1}\cdots\delta_{n_d,m_d}\), hence \(P = 1\).

In Fig.~2(g), we show the participation ratio \(P_1\) of the eigenstate \(\ket{u_1}\), associated with the eigenvalue \(\epsilon_1\). As evident, $P_1$ drops sharply for \(V > 2\), indicating strong localization. Indeed, beyond the threshold, the participation ratio $P_1$ remains constant as the system size increases, i.e., \(P_1 \propto N^0\), as shown in the Supplementary Note 2. This scaling confirms that the isolated eigenvalue corresponds to a localized mode, serving as a clear signature of the infinite-sized lattice. The spatial profile of this eigenstate, depicted in Figs.~3(c) and 3(d) (red curves), exhibits clear exponential localization, similar to bound states induced by Hermitian impurities. 

To investigate the localization properties of the system beyond the isolated bound state, we analyze the participation ratios of eigenstates across different spectral regions as a function of defect strength \(V\), as shown in Fig.~2(g). In the odd-sized lattice (\(N = 51\)), eigenstates associated with purely real eigenvalues remain unaffected by the defect and maintain their unperturbed extended profiles. Of course, their {average} participation ratio, denoted as \(P_R\), remains constant with increasing \(V\) and scales linearly with system size, \(P_R \propto N\), as expected for extended states. Representative examples of such states are shown by the gray curves in Figs.~3(a) and 3(c).

In contrast, the complex-eigenvalue states—excluding the isolated localized state that emerges for \(V>2\)—exhibit a nontrivial dependence on \(V\).  In the odd-sized lattice, this is quantified by the average participation ratio \(P_C^{(2M+1)}\), which is not constant and generally exceeds \(P_R\) for most values of \(V\). A similar trend is observed in the even-sized lattice for the respective metric \(P_C^{(2M)}\) when \(V < V_{\text{th}}\). Although both \(P_C^{(2M)}\) and \(P_C^{(2M+1)}\) scale linearly with \(N\), indicating that these states remain extended in the formal sense, their spatial structure is far from uniform. In fact, many of these eigenstates are peaked around the defect, though extending across the lattice (and thus have large participation ratios), as illustrated by the blue curves in Figs.~3(a)–3(c). These modes exhibit localized profiles with system-size dependent localization length and have been recently termed {scale-free localized} (SFL) in the literature \cite{Li2023, Guo2023}. Their emergence is intrinsically related to the non-Hermitian character of the defect.

Actually, for \(V > V_{\text{th}}\), \(P_C^{(2M)}\) exhibits an abrupt drop, showing a sharp enhancement of localization. In this case, due to the lack of mirror-symmetry, {scattering} states emerge just above the threshold, characterized by asymmetric spatial profiles with larger amplitudes toward one side of the defect, as shown by the green curves in Fig.~3(d). Of course, these states are not unique to non-Hermitian systems; they also appear in non-symmetric Hermitian lattices once a localized mode is induced by a real defect. Finally, we emphasize that not all complex-eigenvalue modes display SFL profiles—as evidenced by the gray curve in Fig.~3(b)—highlighting the need for a more systematic criterion to distinguish SFL modes from extended or scattering states.

To systematically analyze the behavior of the non-Hermiticity-induced SFL states described above, we introduce a metric \(r\), defined as the ratio of the mean amplitude of an eigenmode around the lattice center (where the impurity is placed) and its mean amplitude near the two lattice edges. By convention, for the eigenstate \(\ket{u_j}\) of a 1D tight-binding lattice of \(N\) sites with a defect at site \(m\) we define \(r_{j}\) as
\begin{equation}
    r_j \;=\;\frac{\displaystyle\frac{1}{5}\sum_{n=m-2}^{m+2}\bigl|u_{j,n}\bigr|}
    {\displaystyle\frac{1}{10}\Bigl(\sum_{n=1}^{5}\bigl|u_{j,n}\bigr|
+\sum_{n=N-4}^{N}\bigl|u_{j,n}\bigr|\Bigr)}
\end{equation}
Using this definition, the numerator is the average amplitude over the central lattice region (sites \(m-2\) to \(m+2\)), while the denominator is the average amplitude over the two edge intervals (sites \(1\) to \(5\) and \(N-4\) to \(N\)). By definition, a state can be considered "localized" toward the center only if \(r > 1\); however, in practice, eigenmodes with \(r \lesssim 1.2\) are indistinguishable from extended states. Using this criterion ($r\geq1.2$), we find that eigenstates associated with real eigenvalues in the even-sized lattice (\(N = 50\)) yield a mean ratio of \(r_R^{(50)} = 0.78\), confirming their extended character. In Fig. ~4(a)–(b) we show representative eigenvalue spectra for \(N = 51\) and \(N = 50\), with eigenvalues color-coded according to their localization type; eigenstates with non-real eigenvalues satisfying \(r > 1.2\) are identified as SFL modes (blue), distinguishing them from extended modes (gray), scattering modes (green), and localized modes (red).

It musth be noted that states associated with complex eigenvalues—excluding the isolated eigenvalue that emerges at the threshold—consistently exhibit higher mean \(r\) values, with markedly stronger localization observed in the mirror-symmetric (\(N = 51\)) lattice, as shown in Fig.~4(c). As the defect strength \(V\) approaches the threshold, the mean ratio \(r\) increases, reaching a maximum of approximately \(r_C^{(51)} = 2.6\) for the odd-sized lattice and \(r_C^{(50)} = 1.3\) for the even-sized lattice, before rapidly decreasing for \(V > 2\).
\begin{figure}[H]          
  \centering
  \includegraphics[width=0.5\textwidth]{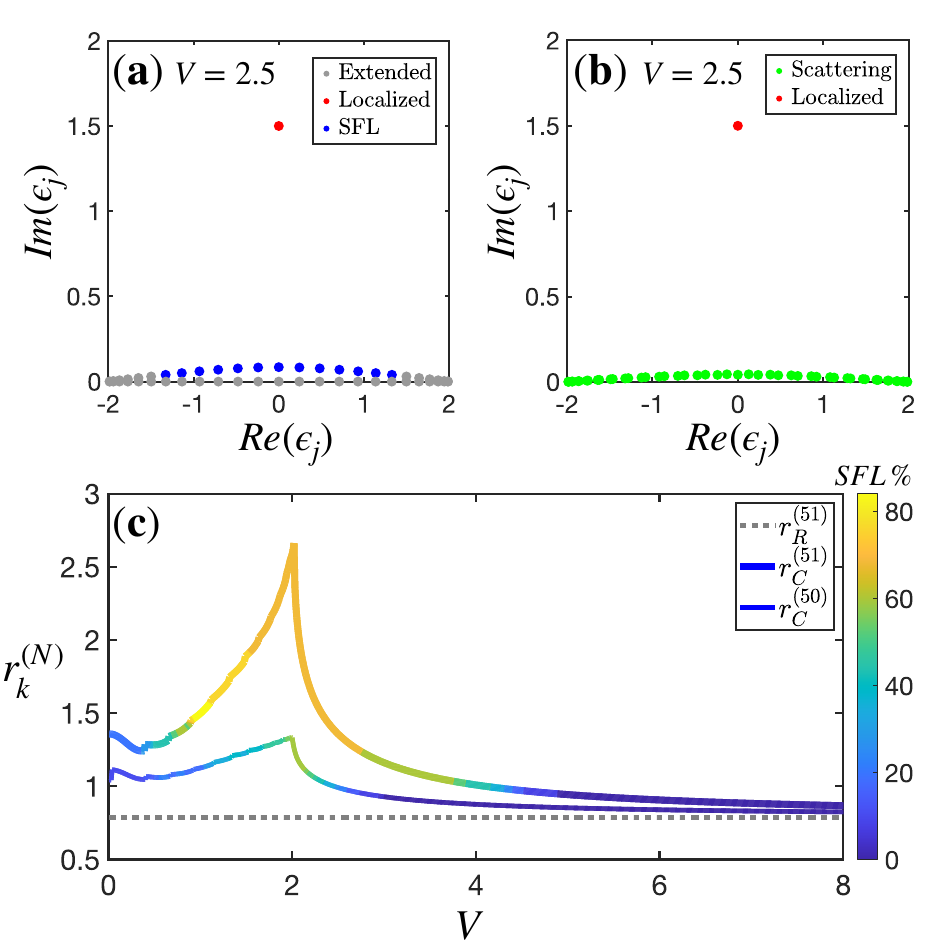}
  \caption{\textbf{Classification of eigenmodes and scale-free localization for 1D finite-sized lattices.}
    (a)/(b) Eigenvalue spectra of 1D tight-binding finite-sized lattices with a single imaginary defect of strength $V=2.5$ for (a) mirror-symmetric lattice (\(N = 51\)) and the defect placed on site $m=26$ and (b) the non-mirror-symmetric lattice (\(N = 50\)) and the defect places on site $m=25$. In both subfigures the eigenvalues are colored according to the classification of their eigenstates; localized (red), SFL (blue), extended (gray) and scattering (green). 
    (c) Ratios $r$ for eigenstates with complex eigenvalues are shown for the mirror-symmetric lattice (\(r_{C}^{(51)}\)) and the non mirror-symmetric lattice (\(r_{C}^{(50)}\)); for comparison, the corresponding ratio for real-eigenvalue states of the mirror-symmetric lattice (\(r_{R}^{(51)}\)) is included (gray line). The color for \(r_{C}^{(51)}\) and \(r_{C}^{(50)}\) indicates the percentage of complex-eigenvalue eigenstates with \(r>1.2\).
  }
  \label{fig:Schematic}
\end{figure}
Importantly, in the odd-sized lattice, more than 70\% of the complex eigenstates satisfy the condition \(r > 1.2\) within the defect strength range \(1.5 \leq V \leq 2.3\), exhibiting SFL behavior. In contrast, the corresponding fraction in the non-mirror-symmetric (\(N = 50\)) lattice peaks at roughly 60\% around the threshold \(V = 2\). For larger values of \(V\), the fraction of states satisfying \(r > 1.2\) declines sharply, vanishing entirely beyond \(V > 4.9\) in the odd lattice and \(V > 2.4\) in the even lattice. These critical values mark the transition points beyond which SFL states are replaced by symmetric extended states (in the mirror-symmetric configuration) and asymmetric scattering states (in the non-symmetric configuration).
We note that adopting PBC yields results qualitatively similar to those of the mirror-symmetric case, as PBC inherently enforce a symmetric geometry (a detailed analysis can be found in the Supplementary Note 3).

{Here, we emphasize that the localized state in the finite-sized lattice corresponds to the bound state that emerges in the infinite-lattice limit, with its eigenvalue \(\epsilon_{1}\) converging to the bound-state value \(\epsilon_{b}\) determined by Eq.~\eqref{pole}. In contrast, as already mentioned, all other eigenstates with non-zero imaginary parts, whose features were discussed, collapse into the unperturbed band as \(N \to \infty\), i.e., \(\langle b \rangle \to 0\), indicating their evolution into extended states.}

Having systematically examined the localization properties induced by a single imaginary defect in 1D lattices, extend our analysis to higher-dimensional systems, specifically 2D square and 3D cubic lattices, described by Eqs.~\eqref{hamiltonian_2d} and \eqref{hamiltonian_3d}, respectively, with OBC imposed along all spatial directions. 
\begin{figure}[H]
  \centering
  \includegraphics[width=0.9\textwidth]{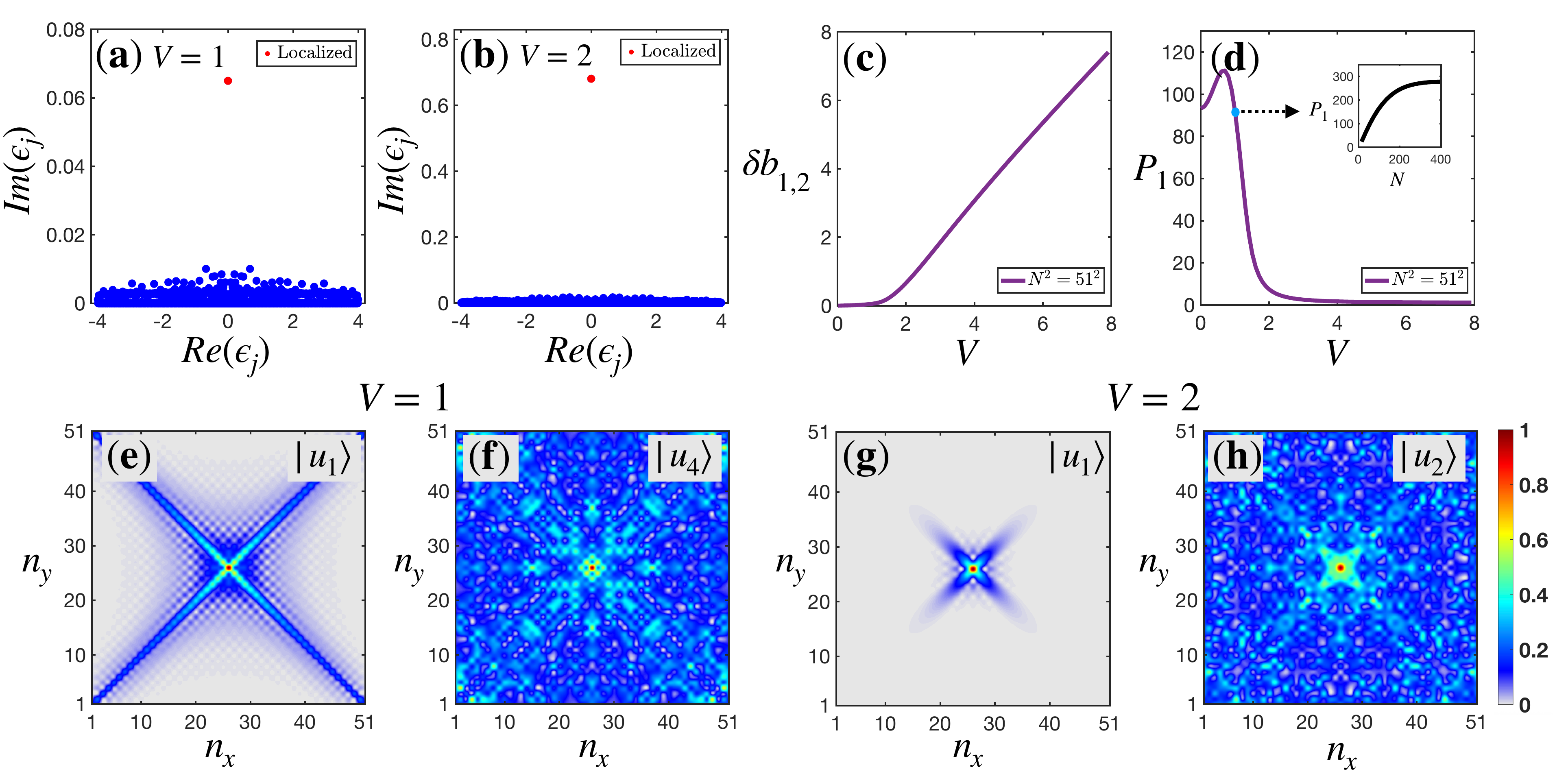}
  \caption{\textbf{Spectral properties and representative eigenmodes of 2D finite-sized lattices with a single imaginary impurity.} 
    (a)–(b) Eigenvalue spectra for 2D finite-sized tight-binding lattices of size \(51\times51\) with a single imaginary defect at \(\ket{\mathbf{m}}=(26,26)\), for (a) \(V=1\) and (b) \(V=2\). The eigenvalue corresponding to a localized state is marked in red.  
    (c) Difference \(\delta b_{1,2}\) as a function of \(V\).  
    (d) Participation ratio \(P_1\) of the eigenstate with the largest imaginary part versus \(V\) (inset: \(P_1\) vs.\ \(N\) at \(V=1\)).  
    (e)-(h) Amplitudes \(|u_{j,(n_x,n_y)}|\) of representative eigenmodes for defect strength  
    (e)-(f) \(V=1\), (g)-(h) \(V=2\). Panels (e)/(g) correspond to localized states, while the (f)/(h) are other complex-eigenvalue states. In all panels, the amplitudes are normalized in order to have maximum value equal to unity. 
  }
  \label{fig:Schematic5}
\end{figure}
Throughout this analysis, we focus on lattices with an odd number of sites per spatial dimension, \(N = 2M + 1\), where \(M \in \mathbb{N}\), placing the defect at the center; \(\ket{\mathbf{m}} = \ket{M+1, M+1}\) in 2D and \(\ket{\mathbf{m}} = \ket{M+1, M+1, M+1}\) in 3D. Unlike the 1D case, where a defect positioned in the center divides the lattice into two distinct segments, the introduction of a defect at the center of a higher-dimensional lattice does not result in such a partitioning. Therefore, the difference between our choice and an even-sized (per dimension) lattice, with the defect at \(\ket{M, M}\) (in 2D) or \(\ket{M, M, M}\) (in 3D), is minimal. Nonetheless, our symmetric configuration aligns more closely with the symmetry of the infinite-lattice limit. Finally, we note that the unperturbed (\(V = 0\)) spectra in both 2D and 3D have eigenvalue degeneracies, and the proportion of eigenvalues that remain purely real for \(V > 0\) depends highly on the system size.

\textbf{2D square lattice:} First, as shown in Figs.~5(a)–(b), the eigenvalue spectra of a 2D lattice with \(N^2 = 51^2\) sites reveal the emergence of a single eigenvalue with a larger imaginary part than the rest of the spectrum, for both \(V = 1\) [Fig.~5(a)] and \(V = 2\) [Fig.~5(b)]. Actually, as illustrated in Fig.~5(c), the difference \(\delta b_{1,2} = \Im(\epsilon_1) - \Im(\epsilon_2)\) is nonzero even for infinitesimally small values of the defect strength \(V\), in sharp contrast to the 1D case. This behavior, which remains robust across different lattice sizes, serves as a clear fingerprint of bound-state formation in 2D, as established in the infinite-lattice section.

The localized character of the eigenmode \(\ket{u_1}\), associated with the eigenvalue having the largest imaginary part \(b_1\), is further confirmed by the behavior of its participation ratio \(P_1\). Unlike the 1D case—where a sharp drop in \(P_1\) indicates a localization threshold—here the participation ratio decreases gradually with increasing \(V\), following a mild peak around \(V \approx 0.7\) [Fig. 5(d)]. For sufficiently large defect strengths (\(V \gtrsim 2\)), \(P_1\) clearly saturates with increasing lattice size, scaling as \(P_1 \propto N^0\), consistent with exponential localization. However, for smaller \(V\), this scaling behavior only becomes apparent in very large lattices. This is demonstrated in the inset of Fig.~5(d), which shows that for \(V = 1\), the localized nature of \(\ket{u_1}\) (i.e., \(P_1 \propto N^0\)) is evident only for lattice sizes exceeding \(N > 400\).

\begin{figure*}
    \centering
    \includegraphics[width=0.45\textwidth]{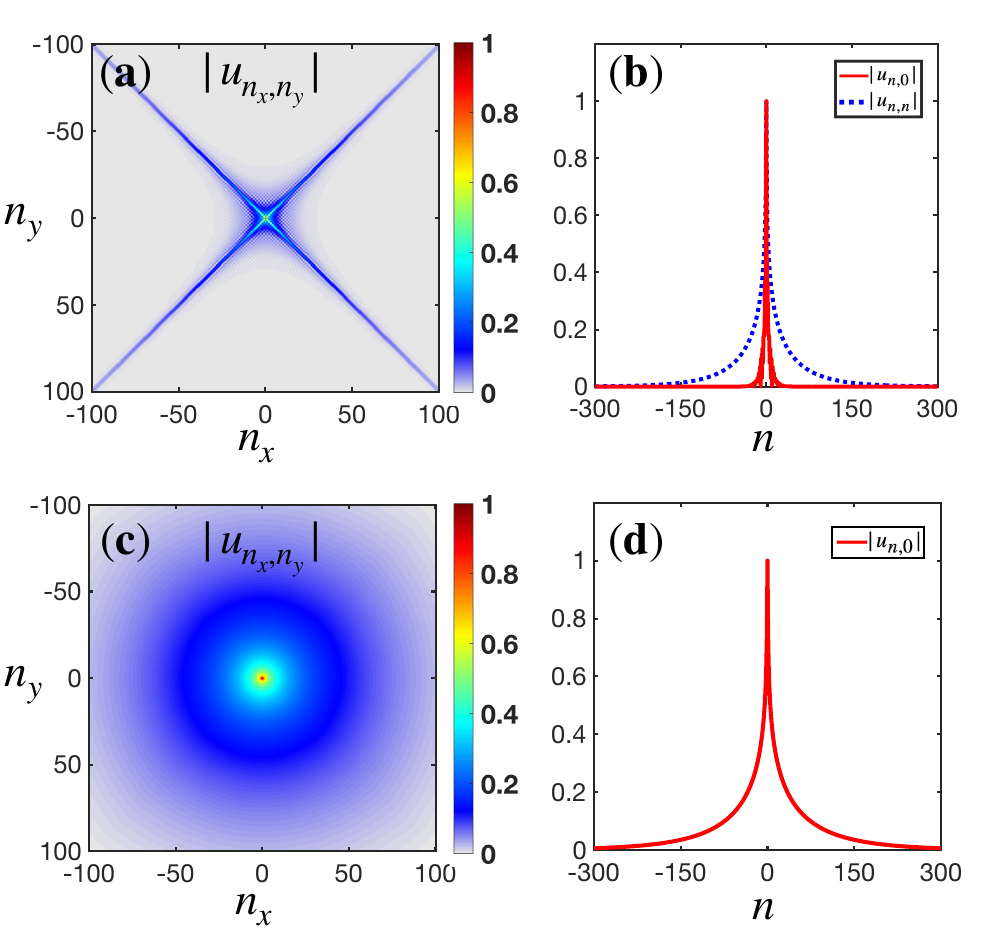}
    \caption{\textbf{Non-Hermitian cross localized states versus conventional exponentially localized states in 2D infinite square lattices.} (a) Amplitude of the non-Hermitian cross-localized (NHCL) eigenstate in a 2D {infinite} lattice for an {imaginary} defect with strength \(V=1\) as a function of the distances \(n_x\) and \(n_y\) from the defect site (set at the origin).  (b) Cross sections of (a) taken along the diagonal \(n_x=n_y=n\) (blue line) and along the horizontal \(n_y=0\) (red line). {For comparison}: (c) Amplitude of the localized eigenstate in a 2D {infinite} lattice for a {real} defect with strength \(V=1\) as a function of the distances \(n_x\) and \(n_y\) from the defect site (set at the origin). (d) Cross section of (c) taken along the horizontal \(n_y=0\) (red line). In all subfigures, the amplitudes are normalized in order to have maximum value equal to unity.}
    \label{fig:Schematic}
\end{figure*}

In the bottom row of Fig. 5 we show representative examples of eigenstates for $V=1$ and $V=2$. In particular, Fig. 5(e) and Fig. 5(g) correspond to the localized states (red-point eigenvalues in Fig. 5(a) and Fig. 5(b) respectively), while Fig. 5(f) and Fig. 5(h) correspond to other complex-eigenvalue states. Importantly, the defect-induced localized state at small values of \(V\) exhibits a nontrivial spatial profile. As shown in Fig.~5(e) for \(V=1\), the amplitude distribution forms a {cross-shaped} pattern with a pronounced central peak. This structure is clearly distinct from the conventional exponential localization typically associated with Hermitian defects. Importantly, this feature is neither a consequence of the applied boundary conditions nor a finite-size artifact—it persists even in the infinite-lattice limit, as verified through Green's function analysis.
In this limit, $N\to \infty$, the bound state corresponding to an eigenvalue \(\epsilon_b\) can be written as
$\ket{u} = \sum_{n_x,n_y=-\infty}^{\infty} u_{n_x,n_y} \ket{n_x,n_y},
$ with $u_{n_x,n_y} \propto G_0\left((n_x,n_y), 0; \epsilon_b\right)$, where \(G_0\left((n_x,n_y), 0; \epsilon_b\right)\) denotes the {off-diagonal} elements of the unperturbed Green’s function \cite{Economou2006}. 
As shown in Fig.~6(a), for an infinite 2D lattice with \(V=1\) (yielding a bound state with eigenvalue \(\epsilon_b \approx 0.029i\)), the cross-shaped amplitude pattern remains clearly visible within a \(200 \times 200\) region around the defect, which is set at the origin. In this region, the decay along the diagonals is notably slower than the rest directions. Indeed, Fig.~6(b) shows that the amplitude \(|u_{n_x,n_y}|\) decreases much more slowly along the diagonal direction (\(n_x = n_y \equiv n\)) than along the horizontal axis (\(n_y = 0\)); along the diagonals, the amplitude only approaches zero for \(n_x > 300\), in contrast to the horizontals where, after some oscillations, vanishes for \(n_x > 40\). Thus, the localized nature of these eigenstates in the infinite-size limit is confirmed, even though their profile does not resemble conventional exponentially localized states. 

To highlight their distinct qualitative differences from Hermitian impurity-induced bound states, we present in Fig.~6(c)-(d) the corresponding amplitude and cross-section for an infinite 2D lattice with a real defect of the same strength, 
$V=1$. Although the eigenstates in Fig.~6(a) and Fig.~6(c) arise from defects of equal magnitude, the non-Hermitian character of the former leads to a fundamentally different structure, compared to the radially symmetric mode of the latter. To the best of our knowledge, such eigenstates—here termed {non-Hermitian cross localized} (NHCL) states—have not been previously reported in the literature. As the defect strength 
$V$ increases, these NHCL eigenstates transition smoothly into conventional exponentially localized modes (see Fig.~5(g) for 
$V=2$). In the Supplementary Note 4, we discuss how such NHCL could be observed dynamically during the propagation of electric field's envelope, in two-dimensional evanescently coupled waveguide lattices, within the context of coupled-mode theory. {Moreover, in the Supplementary Note 5 we have included a detailed mathematical explanation for the emergence of such states.}

\textbf{3D cubic lattice:} In contrast to the 2D square lattice, the eigenvalue spectra of a 3D cubic lattice with \(N^3 = 15^3\) sites exhibit symmetric eigenvalue pairs about the imaginary axis for relatively small defect strengths [e.g., \(V = 1.5\), Fig.~7(a)], with no eigenvalue distinctly separating from the rest of the spectrum. However, at larger defect strengths [e.g., \(V = 2.7\), Fig.~7(b)], a single eigenvalue emerges with a significantly larger imaginary part. Analogously to the 1D case, the difference \(\delta b_{1,2} = \Im(\epsilon_1) - \Im(\epsilon_2)\) becomes nonzero only beyond a finite threshold, identified at \(V_{\text{th}} \approx 2.3\) [Fig.~7(c)]. At this threshold, the previously symmetric pair of eigenvalues coalesces into an {exceptional point}. Here it should be noted that this threshold decreases as the system size increases, converging to the value $V_{\text{th}}=2.23$, which is the threshold in the thermodynamic limit. 
\begin{figure}[H]
  \centering
  \includegraphics[width=0.9\textwidth]{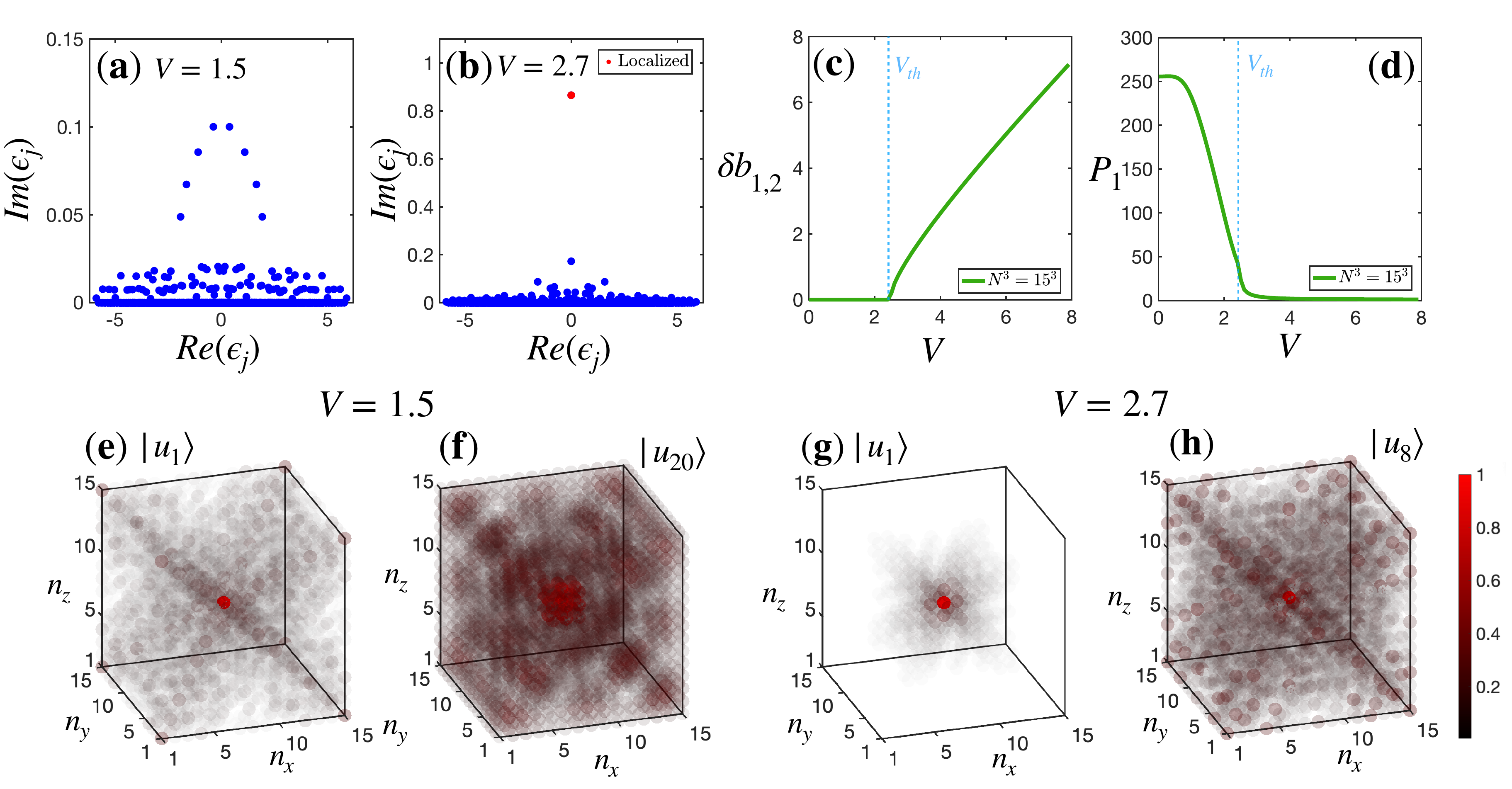}
  \caption{\textbf{Spectral properties and representative eigenmodes of 2D finite-sized lattices with a single imaginary impurity.} 
    (a)–(b) Eigenvalue spectra for 3D finite-sized tight-binding lattices of size \(15\times15\times 15\) with a single imaginary defect at \(\ket{\mathbf{m}}=(8,8,8)\), for (a) \(V=1.5\) and (b) \(V=2.7\). The eigenvalue corresponding to a localized state is marked red.  
    (c) Difference \(\delta b_{1,2}\) as a function of \(V\).  
    (d) Participation ratio \(P_1\) of the eigenstate with the largest imaginary part versus \(V\). In (c) and (d), the blue vertical line marks the threshold value $V_{\text{th}}$ for formation of a localized state.
    (e)-(h) Amplitudes \(|u_{j,(n_x,n_y,n_z)}|\) of representative eigenmodes for defect strength  (e)/(f) \(V=1.5\), (g)/(h) \(V=2.7\). Panel (g) corresponds to a localized state, while the (e)/(f)/(h) are other complex-eigenvalue states. In all subfigures, the amplitudes are normalized in order to have maximum value equal to unity.
  }
  \label{fig:Schematic6}
\end{figure}
Furthermore, as shown in Fig.~7(d), the participation ratio \(P_1\) of the eigenmode \(\ket{u_1}\) decreases monotonically with increasing \(V\), exhibiting a discontinuity in its derivative at the threshold \(V=V_{\text{th}}\). For \(V > V_{\text{th}}\), \(P_1\) remains constant with increasing system size, scaling as \(P_1 \propto N^0\), confirming the localized character of the corresponding eigenstate. {Here we note that, as in 1D, the single localized mode of the finite 2D or 3D lattice evolves into the bound state of the infinite system, with its eigenvalue \(\epsilon_{1}\) converging to the value \(\epsilon_{b}\) determined by Eq.~\eqref{pole}. In contrast, all other modes with nonzero imaginary parts merge back into the unperturbed band as \(N\to\infty\), i.e., \(\langle b\rangle\to0\).}

It is important to note that the emergence of a single eigenvalue with the largest imaginary part generally serves as an indicator, {but not definitive proof}, for the presence of a localized eigenstate. A notable exception occurs in 1D and 3D lattices with \(N = 4q + 1\), \(q \in \mathbb{N}\), where a single eigenvalue with maximum imaginary part appears even for infinitesimally small \(V > 0\); however, in these cases, localization of the associated eigenmode arises only above the corresponding threshold \(V_{\text{th}}\). 

Beyond bound eigenstates, as in the 1D case, several states corresponding to complex eigenvalues exhibit {SFL-like} behavior, in the sense that they have higher amplitudes in the location of the defect than in the rest of the lattice. However, we cannot classify them as SFL states in the formal sense, in terms of the scaling behavior of their localization length. These states are just characterized by a higher amplitude at the defect position while retaining an extended-state scaling behavior \(P\propto N^d\). Representative examples of such eigenstates are shown in Fig.~5 for 2D [Fig.~5(f) and Fig.~5(h)] and 3D lattices, both below [Fig.~7(e) and Fig.~7(f)] and above [Fig.~7(h)] the threshold. Although these eigenmodes display enhanced amplitudes near the lattice center compared to the edges (in 2D) or faces (in 3D), their localization is considerably weaker than that of the localized eigenstates [Fig.~5(g) and Fig.~7(g)].
\subsubsection*{Complex impurity}

Following our detailed analysis of the purely imaginary defect case, we briefly discuss here the case of a complex defect \(\gamma = V_1 + iV_2\), with \(V_1, V_2 >0\).
\begin{figure*}
    \centering
    \includegraphics[width=0.9\textwidth]{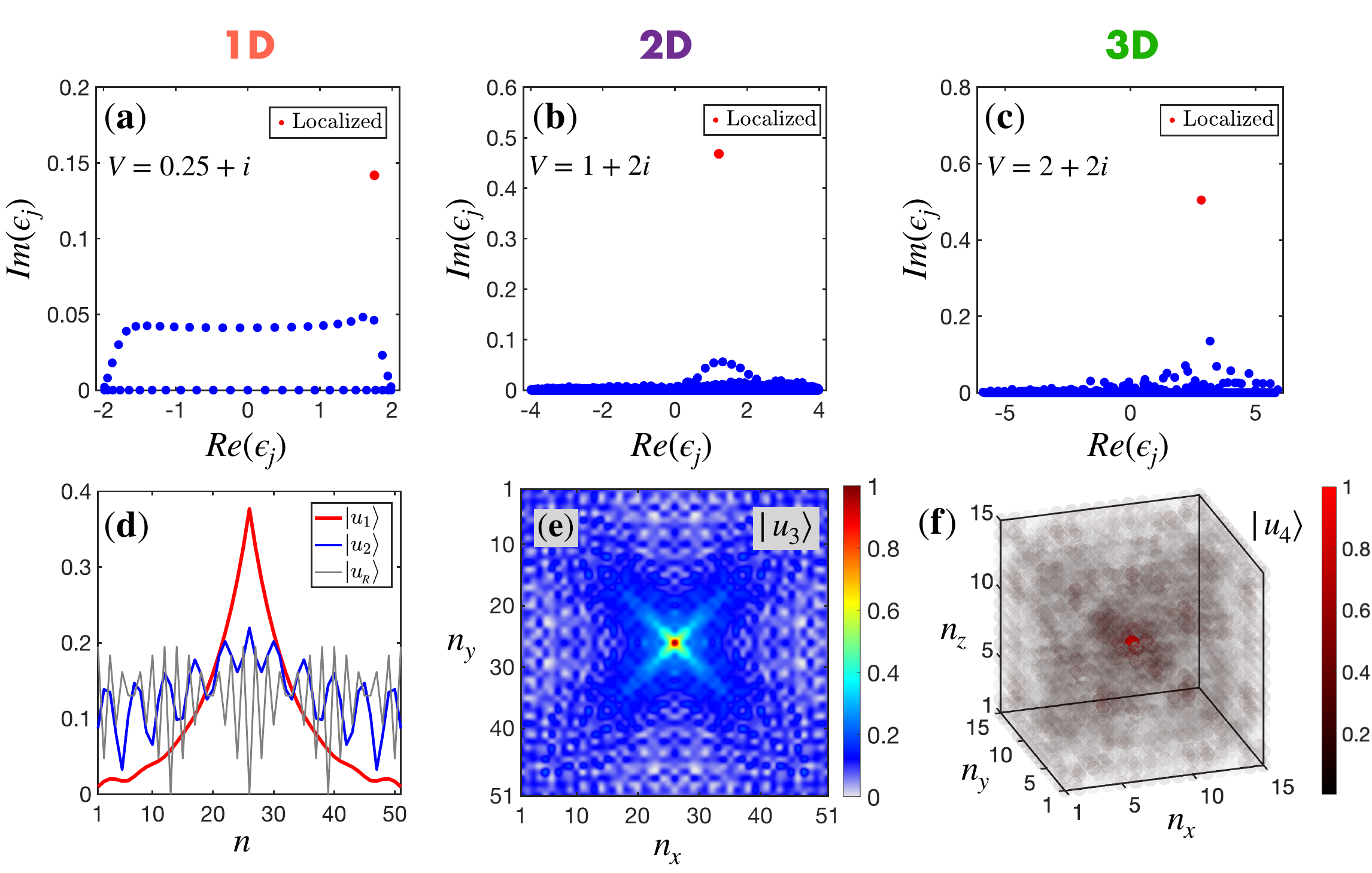}
    \caption{\textbf{Representative eigenvalue spectra and eigenmode amplitudes for finite-sized lattices in 1D, 2D and 3D with a single complex impurity.} (a) Eigenvalue spectrum for a 1D {finite-sized} tight-binding lattices of size \(N=51\) with a single complex defect $V=0.25+i
$ located at $m=26$. (b) Eigenvalue spectrum for a 2D {finite-sized} tight-binding lattice of size \(51\times 51\) with a single complex defect $V=1+2i$ located at \(\ket{\mathbf{m}}=(26,26)\). (c) Eigenvalue spectrum for a 3D {finite-sized} tight-binding lattice of size \(15\times 15\times 15\) with a single complex defect $V=2+2i$ located at \(\ket{\mathbf{m}}=(8,8,8)\). For all panels of top row, the eigenvalue corresponding to the localized mode is marked with red color.
    (d) Amplitudes  of representative eigenmodes \(|u_{j,n}|\)   corresponding to localized (red), SFL (blue) and extended (gray) states for the 1D case of (a). In particular, the extended real-eigenvalue eigenmode (gray) $\ket{u_R}$ of (a) corresponds to ${u_{R,n}}\propto \sin(\frac{7\pi n}{13})$ (e)/(f) Amplitudes of pertinent eigenmodes displaying {SFL-like} profiles for the 2D/3D cases of (b)/(c), respectively. In panels (e)/(f), the amplitudes are normalized in order to have maximum value equal to unity.}

    \label{fig:Schematic}
\end{figure*}

Across lattices of all dimensions (1D, 2D square, and 3D cubic), the presence of a real part in the defect's strength breaks the spectral symmetry with respect to the imaginary axis, a symmetry characteristic of the purely imaginary defect case discussed extensively in the previous paragraph.
Consequently, as illustrated by representative eigenvalue spectra in the top row of Fig.~8, the mean real part of the eigenvalues \(\epsilon_j=a_j+ib_j\) becomes nonzero, while many eigenvalues acquire nonzero imaginary parts \(b_j\), depending on dimensionality, lattice size, and impurity location. As detailed in the first part of this work, the existence or absence of a localized eigenstate depends on the lattice dimensionality and the defect strength's real and imaginar parts \(V_1\) and \(V_2\). Note that in all cases presented in Fig.~8, these parameters are chosen such that a localized state exists. However, due to the broken symmetry with respect to the imaginary axis, no clear relationship emerges between the existence of nonzero \(\delta b_{1,2}\) and the localization properties of the eigenmode \(\ket{u_1}\). Thus, generally, an eigenvalue possessing the maximum imaginary part may correspond to an extended state.

Meanwhile, the mean imaginary part \(\langle b \rangle\) of the eigenvalues (excluding the localized state, when present) typically decreases with increasing system size \(N\). In the thermodynamic limit, all eigenvalues remain inside the unperturbed band, except possibly a single bound-state eigenvalue that exists for specific values of \(V_1\) and \(V_2\), as detailed for the infinite lattice in the first section of this work (see Fig.~1).

Regarding the eigenmodes, in 1D, the guaranteed formation of a localized state for $V_2>0$—as established in the first section—results in properties of the remaining modes closely mirroring those of a purely real defect. In lattices lacking mirror symmetry, non-localized modes behave as scattering states, similar to those presented in Fig.~3(d). In mirror-symmetric lattices, these modes appear extended, with possibly a few exhibiting weakly SFL profiles characterized by slightly elevated amplitudes near the lattice center relative to its edges, as illustrated in Fig.~8(d). In 2D and 3D, aside from the localized state (when present), many of the remaining complex eigenmodes exhibit {SFL-like} profiles peaked at the impurity site, examples of which are shown in Figs.~8(e) and 8(f) for 2D and 3D, respectively, while the rest of the spectrum consists of extended states.

\section*{Conclusions}
{In this work, we have addressed the single complex impurity problem, extending a central problem of condensed-matter physics to the non-Hermitian regime.
In summary, we have explored the effects of a single non-Hermitian impurity in one-dimensional (1D), two-dimensional square (2D), and three-dimensional cubic (3D) tight-binding lattices.}

{Starting from the thermodynamic limit, we demonstrated—using Green’s function techniques—that the complex impurity problem is far from a straightforward extension of the Hermitian case. In 1D, we identified threshold emergence for bound-state formation in the presence of a purely imaginary defect. More intriguingly, in 2D and 3D systems, we uncovered the emergence of forbidden regions in the parameter space of the impurity strength's real and imaginary parts, where the coexistence of real and imaginary components suppresses bound-state formation—even though either component alone would support it. It is well known that, in Hermitian systems, the problem of Anderson localization is closely connected to the emergence of a bound state induced by a single impurity. A similar relation is expected to hold in the non-Hermitian case—particularly for complex impurities—linking our findings to the broader question of Anderson localization in systems with complex diagonal disorder. 
Thus, the unexpected behavior regarding bound-state formation reported in our work is particularly relevant in the context of non-Hermitian disordered systems, where it may prompt further investigation—especially toward developing a more intuitive physical understanding of the modified thresholds for Anderson transitions in 3D lattices with complex disorder. Since the single impurity serves as the fundamental building block for disordered lattices, such insights could complement and enrich existing scaling-based approaches \cite{Huang2020, Kawabata2021, Luo2021, Luo2021_b, Luo2022 }.}

Regarding the finite-sized systems, which are more relevant to experimental realizations {in photonics}, we analyzed the resulting spectral and localization properties. Our results reveal a diverse range of non-Hermitian phenomena, including the emergence of scale-free localized states, the formation of exceptional points, and the appearance of cross-shaped localized modes—features that differ qualitatively from those associated with conventional Hermitian impurities. These findings deepen our understanding of impurity effects in non-Hermitian lattices and may guide future experimental investigation across photonics and related platforms.

\section*{Methods}
For the analysis of infinite lattices, the unperturbed Green’s functions were evaluated numerically in MATLAB using standard numerical integration routines. For finite lattices, all eigenvalues and eigenvectors were computed using MATLAB’s built-in diagonalization functions.

\section*{Data availability}
The data supporting the findings of this study are available from the authors upon reasonable request.

\section*{Code availability}
The MATLAB codes used to perform the numerical analyses reported in this work are available from the authors upon reasonable request.

\section*{Acknowledgements}

The authors acknowledge financial support from the European Research Council (ERC) through the Consolidator Grant Agreement No. 101045135 (Beyond Anderson). 

\section*{Author contributions}

E.T.K. and I.K. performed the numerical simulations and analyzed the results. E.N.E. and K.G.M. conceived the idea and supervised the project. All authors contributed to writing the manuscript.

\section*{Competing interests}
The authors declare no competing interests.
\end{document}